\documentclass[11pt]{article}
\usepackage{graphicx}
\usepackage{amsmath}

\newcommand{\BABARPubYear}    {04}

\newcommand{\BABARConfNumber} {039}
\newcommand{\SLACPubNumber} {10655}

\input pubboard/babarsym
 
\long\def\inst#1{\par\nobreak\kern 4pt\nobreak
  {\it #1}\par\vskip 10pt plus 3pt minus 3pt}

\def\bmtodcppk {\ensuremath {B^-{\to}D^0_{\CP+} K^-}}
\def\bptodcppk {\ensuremath {B^+{\to}D^0_{\CP+} K^+}}
\def\btodcppp {\ensuremath {B^-{\to}D^0_{\CP+} \pi^-}}
\def\bmtodcpmk {\ensuremath {B^-{\to}D^0_{\CP-} K^-}}
\def\bptodcpmk {\ensuremath {B^+{\to}D^0_{\CP-} K^+}}
\def\btodcpmp {\ensuremath {B^-{\to}D^0_{\CP-} \pi^-}}
\def\btodk   {\ensuremath {B^-{\to}D^0 K^-}}

\def\btodp   {\ensuremath {B^-{\ra}D^0 \pi^-}}

\def\btodh   {\ensuremath {B^-{\to}D^0 h^-}}
\def\CP                {\ensuremath{C\!P}\xspace}
\def\CPp                {\ensuremath{C\!P\!+}\xspace}
\def\CPm                {\ensuremath{C\!P\!-}\xspace}

\def\kpi{\ensuremath{K^-\pi^+}\xspace}
\def\kk{\ensuremath{K^-K^+}\xspace}
\def\pipi{\ensuremath{\pi^-\pi^+}\xspace}
\def\kspi0{\ensuremath{\KS \piz}\xspace}

\begin{document}
  {\pagestyle{empty}

    \begin{flushright}
      \babar-CONF-\BABARPubYear/\BABARConfNumber \\
      SLAC-PUB-\SLACPubNumber \\
    \end{flushright}
    
    \par\vskip 5cm
    
    \begin{center}
      \Large \bf 
      Measurement of the Branching Fractions and \CP Asymmetries of
      $B^-\ra D^0_{(\CP)}K^-$ Decays with the \babar\ Detector
    \end{center}
    \bigskip
    
    \begin{center}
      \large The \babar\ Collaboration\\
      \mbox{ } \\
      \today
    \end{center}
    \bigskip \bigskip
    
    \begin{center}
      \large \bf Abstract
    \end{center}

    We present a study of $\Bm\to\Dz_{(\CP)}\Km$ decays, where $D^0_{(\CP)}$ is
    reconstructed in flavor (\kpi), \CP-even (\kk,\pipi) and
    \CP-odd (\kspi0) eigenstates,
    based on a sample of about 214 million $\Upsilon(4S)\rightarrow
    \BB$ decays collected with the \babar\ detector at the PEP-II
    \epem storage ring.
    Along with the Cabibbo-suppressed $\Bm\to\Dz_{(\CP)}\Km$ decays we
    reconstruct also the Cabibbo-favored $\Bm\to\Dz_{(\CP)}\pim$
    decays. We measure the double ratio of branching fractions
    \begin{eqnarray*}
      R_{+} & \equiv &
      \frac{\BR(\bmtodcppk)/\BR(\btodcppp)}{\BR(\btodk)/\BR(\btodp)} \\
       & = & 0.87\pm 0.14\stat\pm 0.06\syst,\nonumber
    \end{eqnarray*}
    \begin{eqnarray*}
      R_{-} & \equiv &
      \frac{\BR(\bmtodcpmk)/\BR(\btodcpmp)}{\BR(\btodk)/\BR(\btodp)} \\
      & = & 0.80\pm 0.14\stat\pm 0.08\syst, \nonumber 
    \end{eqnarray*}
    and the \CP asymmetries
    \begin{eqnarray*}
      A_{\CPp} & \equiv &
      \frac{\BR(\bmtodcppk)-\BR(\bptodcppk)}{\BR(\bmtodcppk)+\BR(\bptodcppk)}
      \\ 
      & = & 0.40\pm 0.15\stat\pm 0.08\syst \nonumber
    \end{eqnarray*}
    \begin{eqnarray*}
      A_{\CPm} & \equiv &
      \frac{\BR(\bmtodcpmk)-\BR(\bptodcpmk)}{\BR(\bmtodcpmk)+\BR(\bptodcpmk)}
      \\
      & = & 0.21\pm 0.17\stat\pm 0.07\syst. \nonumber
    \end{eqnarray*}
    All results are preliminary. 

    \vfill
    \begin{center}

      Submitted to the 32$^{\rm nd}$ International Conference on High-Energy Physics, ICHEP 04,\\
      16 August---22 August 2004, Beijing, China

    \end{center}

    \vspace{1.0cm}
    \begin{center}
      {\em Stanford Linear Accelerator Center, Stanford University,  
	Stanford, CA 94309} \\ \vspace{0.1cm}\hrule\vspace{0.1cm}
      Work supported in part by Department of Energy contract DE-AC03-76SF00515.
    \end{center}

    \newpage
  }

  %
  %
  \begin{center}
\small

The \babar\ Collaboration,
\bigskip

%
B.~Aubert,
R.~Barate,
D.~Boutigny,
F.~Couderc,
J.-M.~Gaillard,
A.~Hicheur,
Y.~Karyotakis,
J.~P.~Lees,
V.~Tisserand,
A.~Zghiche
\inst{Laboratoire de Physique des Particules, F-74941 Annecy-le-Vieux, France }
% ...

\end{center}\newpage

  \section{INTRODUCTION}
  \label{sec:Introduction}
  A theoretically clean measurement of the
  angle $\gamma=\arg(-V_{ud}V_{ub}^*/V_{cd}V_{cb}^*)$ can  
  be obtained from the study of $B^-{\ra}D^{(*)0}K^{(*)-}$ decays
  by exploiting the interference between the $b\ra c\bar{u}s$ and
  $b\ra u\bar{c}s$ decay amplitudes~\cite{gronau1991}. 
  The method originally proposed by Gronau, Wyler and London is based on the
  interference between $B^-\ra D^0K^-$ and $B^-\ra \Dzb K^-$ when the
  $D^0$ and \Dzb decay to \CP eigenstates.

  We define the ratios $R$ and $R_{\CP\pm}$ of Cabibbo-suppressed to
  Cabibbo-favored branching fractions 
  \begin{equation}
    R_{(\CP\pm)} \equiv \frac{\BR(B^-{\ra}\Dz_{(\CP\pm)}K^-)+\BR(B^+{\ra}\Dzb_{(\CP\pm)}K^+)}{\BR(B^-{\ra}\Dz_{(\CP\pm)}\pi^-)+\BR(B^+{\ra}\Dzb_{(\CP\pm)}\pi^+)}
  \end{equation}
  with the neutral $D$ meson reconstructed in non-\CP (\Dz) or
  \CP-even/odd eigenstates ($D^0_{\CP\pm}$) channels, 
  and the direct \CP asymmetry
  \begin{equation}
    A_{\CP\pm} \equiv \frac{\BR(B^-{\ra}\Dz_{\CP\pm}K^-)-\BR(B^+{\ra}\Dz_{\CP\pm}K^+)}{\BR(B^-{\ra}\Dz_{\CP\pm}K^-)+\BR(B^+{\ra}\Dz_{\CP\pm}K^+)}\ .
  \end{equation}
  Neglecting the $D^0-\Dzb$
  mixing and the ratio $r_\pi = A(B^-\ra \Dzb\pi^-)/A(B^-\ra
  \Dz\pi^-)$ of the amplitudes of the 
  $B^-\ra \Dzb\pi^-$ and $B^-\ra D^0\pi^-$ processes ($|r_\pi|<
  0.02$), we can write 
  $R_{\pm} \equiv R_{\CP\pm}/R=1+r^2\pm 2r\cos\delta\cos\gamma$ and
  $A_{\CP\pm}=\pm 2r\sin\delta\sin\gamma/(1+r^2\pm
  2r\cos\delta\cos\gamma)$. Here $r = |A(B^-\ra \Dzb K^-)/A(B^-\ra \Dz
  K^-)|$ is the magnitude
  of the ratio of the amplitudes for the processes $B^-\ra \Dzb K^-$
  and $B^-\ra D^0 K^-$, expected from theory to be about 0.1 -- 0.2, and 
  $\delta$ is the relative strong phase between these two
  amplitudes~\cite{gronau1991}. The measurement of $R_{\pm}$ and 
  $A_{\CP\pm}$ allows one to constrain the three unknowns $r$, $\delta$ and 
  the CKM angle $\gamma$.
  In this paper we present the measurement of $R_{\pm}$ and $A_{\CP\pm}$.

  \section{THE \babar\ DETECTOR AND DATASET}
  \label{sec:babar}
  The measurements reported in this paper 
  have been obtained from a sample of about 214 million
  \FourS\ decays to $B\overline{B}$ pairs collected with the 
  \babar\ detector at the \pep2\ asymmetric-energy $B$ factory. 
  The \babar\ detector is described in detail
  elsewhere~\cite{detector}.
  Charged-particle tracking is provided by a five-layer silicon
  vertex tracker (SVT) and a 40-layer drift chamber (DCH). 
  For charged-particle identification, ionization energy loss in
  the DCH and SVT, and Cherenkov radiation detected in a ring-imaging
  device (DIRC) are used. 
  Photons are identified 
  by the electromagnetic calorimeter
  (EMC), which comprises 6580 thallium-doped CsI crystals. 
  These systems are mounted inside a 1.5-T solenoidal
  superconducting magnet. 
  The segmented flux return, including endcaps, is instrumented with
  resistive plate chambers (IFR) for muon and \KL identification.
  We use the GEANT~\cite{geant} software to simulate interactions of particles
  traversing the detector, taking into account the varying
  accelerator and detector conditions. 

  \section{ANALYSIS METHOD}
  \label{sec:Analysis}
  We reconstruct \btodh\ decays, where the prompt track $h^-$ is a kaon
  or a pion. Reference to the charge-conjugate state is implied here and
  throughout the text unless otherwise stated. Candidates for $D^0$ are
  reconstructed in the \CP-even eigenstates $\pi^-\pi^+$ and $K^-K^+$, in
  the \CP-odd eigenstate $K^0_S\pi^0$, and
  in the non-\CP flavor eigenstate $K^-\pi^+$. \KS candidates are
  selected in the \pipi channel.

  The prompt particle $h^-$ is required to have momentum greater than
  1.4 \gevc. 
  Particle ID information from the drift chamber and, when
  available, from the DIRC must be consistent with the kaon
  hypothesis for the $K$ meson candidate in all \Dz\ modes and with the pion 
  hypothesis for the $\pi^\pm$ meson candidates in the $D^0{\ra}\pi^-\pi^+$
  mode.
  For the prompt track to be identified as a pion or a kaon,
  we require that at least five Cherenkov photons are detected 
  to insure a good measurement of the Cherenkov angle.
  We reject a candidate track
  if its Cherenkov angle is not within 3$\sigma$ of the expected value
  for either the kaon or pion mass hypothesis. We also reject
  candidate tracks that are identified as 
  electrons by the DCH and the EMC or as muons by the DCH and the IFR.

  Photon candidates are clusters in the EMC that are not matched to any
  charged track, have a raw energy greater than 30 \mev and lateral
  shower shape consistent with the expected pattern of energy deposit
  from an electromagnetic shower.
  Photon pairs with invariant mass within the range 115--150 \mevcc 
  ($\sim$3$\sigma$) and
  total energy greater than 200 \mev are considered \piz candidates.
  To improve the momentum resolution, the $\piz$ candidates are kinematically
  fit with their mass constrained to the nominal \piz\ mass~\cite{PDG2004}.

  Neutral kaons are reconstructed from pairs of oppositely charged
  tracks with the 
  invariant mass within 10~\mev ($\sim$3$\sigma$)
  from the nominal $K^0$ mass. We also require
  that the ratio between the flight length distance in the plane transverse to 
  the beams direction and its uncertainty is greater than 3.

  The invariant mass of a \Dz candidate, $m(D^0)$, must be
  within 3$\sigma$ of the \Dz mass. The \Dz mass resolution $\sigma$
  is about 7.5~\mev in the \kpi, \kk and \pipi modes, and about
  21~\mev in the \kspi0 mode.
  Selected \Dz candidates are fitted with a constraint to the
  nominal \Dz mass. 

  We reconstruct $B$ meson candidates by combining a \Dz\ candidate
  with a track $h^-$. For the $K^-\pi^+$ mode, the charge of the
  track $h^-$ must match that of the kaon from the $D^0$ meson decay.
  We select $B$ meson candidates by using the beam-energy-substituted mass 
  $\mes = \sqrt{(E_i^{*2}/2 + \mathbf{p}_i\cdot\mathbf{p}_B)^2/E_i^2-p_B^2}$
  and the energy difference $\Delta E=E^*_B-E_i^*/2$, 
  where the subscripts $i$ and $B$ refer to the initial \epem\ system and the 
  $B$ candidate respectively, and the asterisk denotes the
  center-of-mass (CM)
  ($\Upsilon(4S)$) frame.  
  The \mes\ distributions for \btodh\ signal events are Gaussian
  distributions centered at the $B$ mass with a resolution of $2.6
  \mevcc$, which does not depend on the decay mode or on the nature of
  the prompt track.
  In contrast, the \DeltaE\ distributions depend on the mass assigned to the
  prompt track. We evaluate $\Delta E$ with the kaon mass hypothesis 
  so that the distributions are centered near zero
  for \btodk\ events and shifted, on average, by approximately $50
  \mev$ for \btodp\ events. 
  The \DeltaE\ resolution depends on the momentum resolution for the
  \Dz meson and the prompt track $h^-$, and is typically $17\mev$
  for all the \Dz\ decay modes. All $B$ candidates are selected 
  with \mes within 2.5$\sigma$ of the mean value and with \DeltaE in the 
  range $-0.15<\Delta E<0.18\gev$. 

  To reduce background from continuum production of light quarks, we construct
  a Fisher discriminant based on the following quantities: (i) the scalar
  sum of the momenta of all charged and neutral particles (exluding
  the $B$ decay products) flowing into nine concentric cones centered
  on the $B$ candidate thrust axis in the CM frame; 
  (ii) the normalized second Fox-Wolfram moment~\cite{fox_wol}, 
  $R_2\equiv H_2/H_0$, where $H_l$ is the $l$--order Fox-Wolfram moment 
  of all the charged tracks and neutral clusters in the event; 
  (iii) $|\cos\theta_T|$, where $\theta_T$ is the angle between 
  the thrust axes of the $B$ candidate and of the remaining charged tracks 
  and neutral clusters, evaluated in the CM frame; (iv)
  $|\cos\theta_B|$, where $\theta_B$ is the polar angle of the $B$
  candidate in the CM frame; 
  (v) $|\cos\theta_{hel}(D^0)|$, where $\theta_{hel}(D^0)$ is the angle
  between the direction of one of the decay products of the $D^0$ and the 
  direction of flight of the $B$, in the $D^0$ rest frame. Each cone
  in (i) subtends an angle of $10^\circ$ in the CM and is folded to
  combine the forward and the backward intervals.
  A cut on the value of the Fisher discriminant rejects more than
  $90\%$ of the continuum background while retaining $77\%$ of the
  signal in the \kpi, \kk and \kspi0 modes and $65\%$ in the \pipi channel.

  Multiple \btodh candidates are found  in about $4\%$ of 
  the events for the $K^0_S\pi^0$ and in less than 1\% of 
  the events for the other \Dz\ decays.  In these events a $\chi^2$ 
  is constructed from  $m(\pi^0)$ (for $K^0_S\pi^0$ only),  $m(D^0)$,
  and \mes\ and only the candidate with the smallest $\chi^2$ is retained.
  The total reconstruction efficiencies, based on simulated 
  signal events, are about 33\%($K^-\pi^+$), 28\%($K^-K^+$),
  26\%($\pi^-\pi^+$) and 17\%($K^0_S\pi^0$).

  The main contributions to the \BB background come
  from the processes $B{\ra}D^{*}h$ ($h=\pi,K$), $B^-{\ra}D^0\rho^-$ and
  mis-reconstructed \btodh.
  For $D^0\ra K^-K^+$, $D^0\ra \pi^-\pi^+$ and $D^0\ra K^0_S \pi^0$
  decays, the peaking backgrounds 
  $B^-{\ra}K^-K^+K^-$, $B^-{\ra}K^-\pi^+\pi^-$ and $B^-\ra K^0_S \pi^0
  K^-$ must also be considered, since they have 
  the same \DeltaE\ and \mes\ distribution as the $D^0 K^-$
  signal. Their yields
  are estimated from the existing measurements\cite{PDG2004,btokkk}
  and subtracted from the \btodk\ signal yields.

  For each \Dz\ decay mode an extended unbinned maximum-likelihood fit
  to the selected data events determines the signal and background yields $n_i$
  ($i=1$ to $M$, where $M$ is the total number of signal and background 
  channels). Two kinds of signal events, \btodp\ and \btodk, are considered,
  and four kinds of backgrounds: candidates selected either from
  continuum or from \BB\ events, in which the prompt track is either
  a pion or a kaon. 

  The input variables to the fit are \DeltaE\ and a particle identification 
  probability for the prompt track based
  on the Cherenkov angle $\theta_C$, the momentum $p$ and
  the polar angle $\theta$ of the track.
  The extended likelihood function $\cal L$ is defined as
  \begin{equation}
    {\cal L}= \exp\left(-\sum_{i=1}^M n_i\right)\, \prod_{j=1}^N
    \left[\sum_{i=1}^M n_i {\cal P}_i\left(\DeltaE, \theta_C;
      \vec{\alpha}_i\right) \right]\,,
  \end{equation}
  where $N$ is the total number of observed events. 
  The  $M$ functions ${\cal P}_i(\DeltaE, \theta_C;\vec{\alpha}_i)$ are the
  probability density functions (PDFs) for the variables
  $\DeltaE, \theta_C$, given the set of parameters
  $\vec{\alpha}_i$. Since these two quantities are sufficiently
  uncorrelated, their probability density functions are evaluated as a product
  $\mathcal{P}_i=\mathcal{P}_i(\DeltaE;\vec{\alpha}_i)\times\mathcal{P}_i({\theta_C;\vec{\alpha}_i})$.

  The \DeltaE\ distribution for \btodk\ signal events is parametrized
  with a Gaussian function.
  The \DeltaE\ distribution for \btodp\ is parametrized with the same
  Gaussian used for \btodk\ with a relative shift of the mean, computed event
  by event as a function of the prompt track momentum, arising from the
  wrong mass assignment to the prompt track. The offset and width of
  the Gaussian are kept floating in the fit and are determined from data
  together with the yields.
  
  The \DeltaE\ distribution for the continuum background is parametrized with
  a linear function whose slope is determined from off-resonance data.
  The \DeltaE\ distribution for the \BB background is empirically
  parametrized with the sum of a Gaussian and an exponential function when
  the prompt track is a pion, and with an exponential function when the
  prompt track is a kaon. The parameters are determined from simulated
  events.

  The particle identification PDF is obtained 
  from a pure control sample of kaons and pions produced in the decay
  chain $D^{*+}\to\Dz\pip$ ($\Dz{\to}\Km\pip$), selected using
  kinematical information only, without any 
  inputs from the \babar\ particle identification system.
  The parametrization of the particle identification PDF is performed 
  by fitting with a Gaussian distribution the background-subtracted
  distribution of the difference between the reconstructed and expected
  Cherenkov angles of the selected kaons and pions.

  \section{PHYSICS RESULTS AND SYSTEMATIC STUDIES}
  \label{sec:Systematics}

\begin{figure}[!htb]
    \begin{center}
      \includegraphics[width=9.3cm]{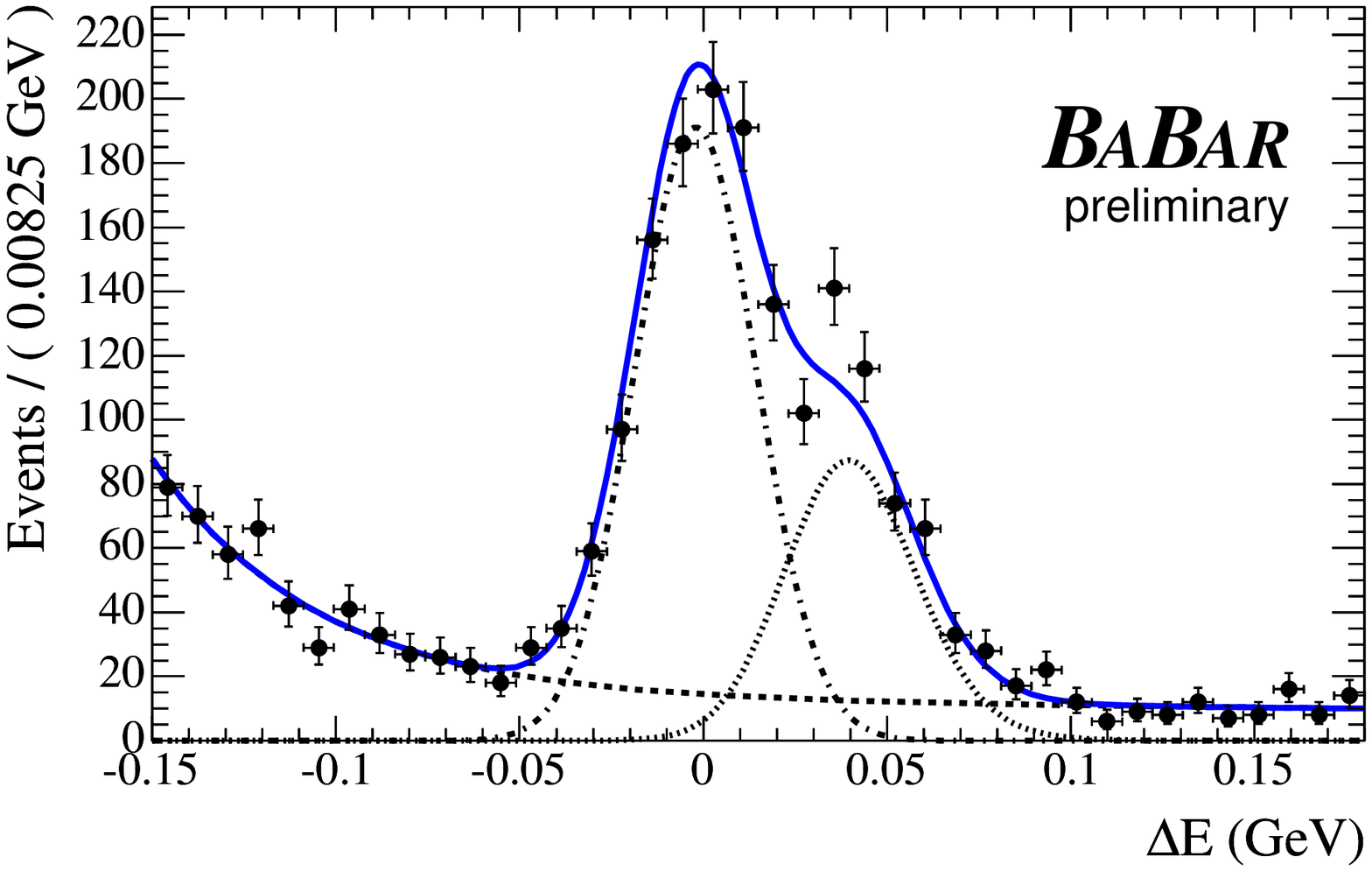}
      \includegraphics[width=9.3cm]{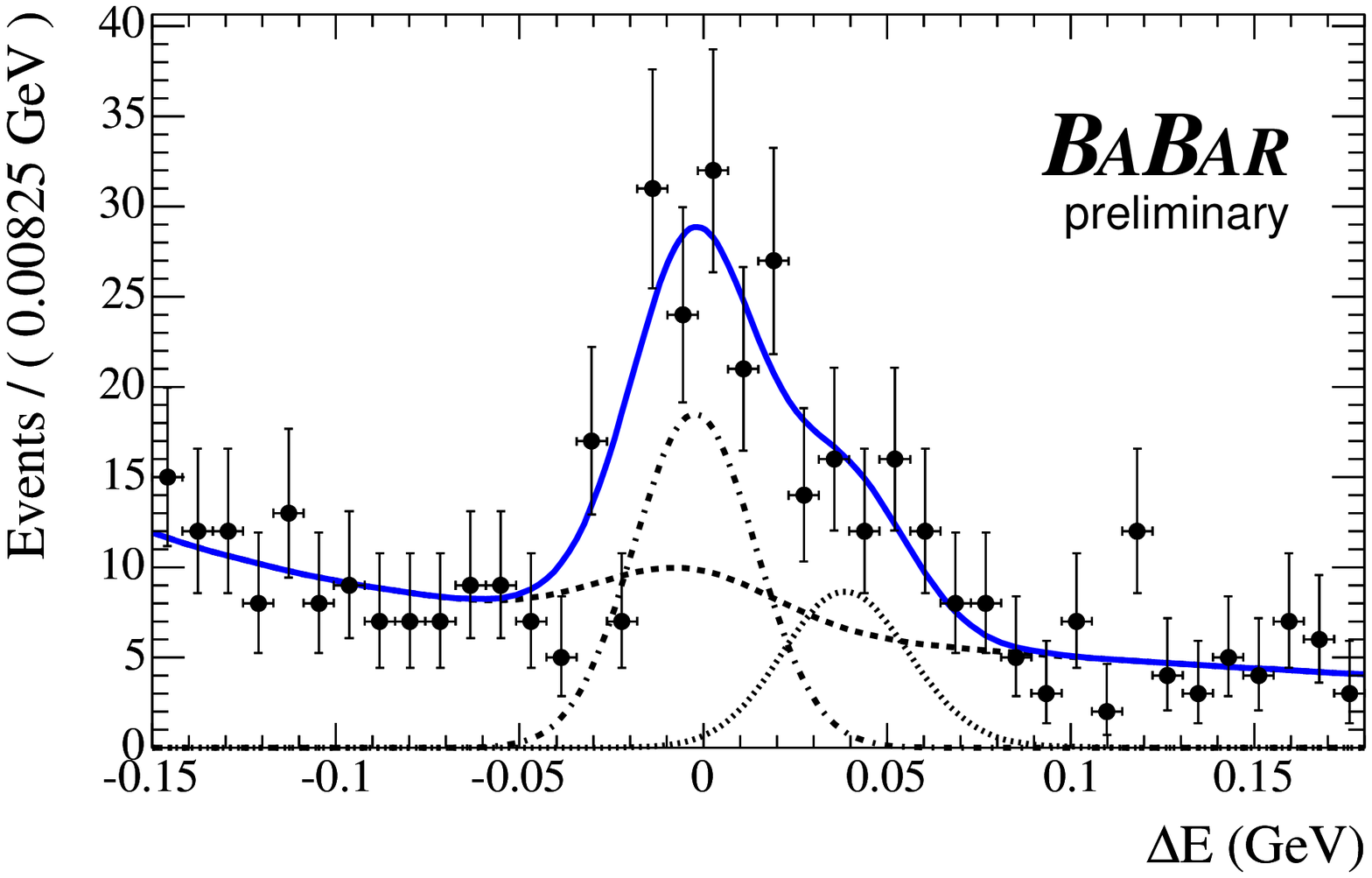}
      \includegraphics[width=9.3cm]{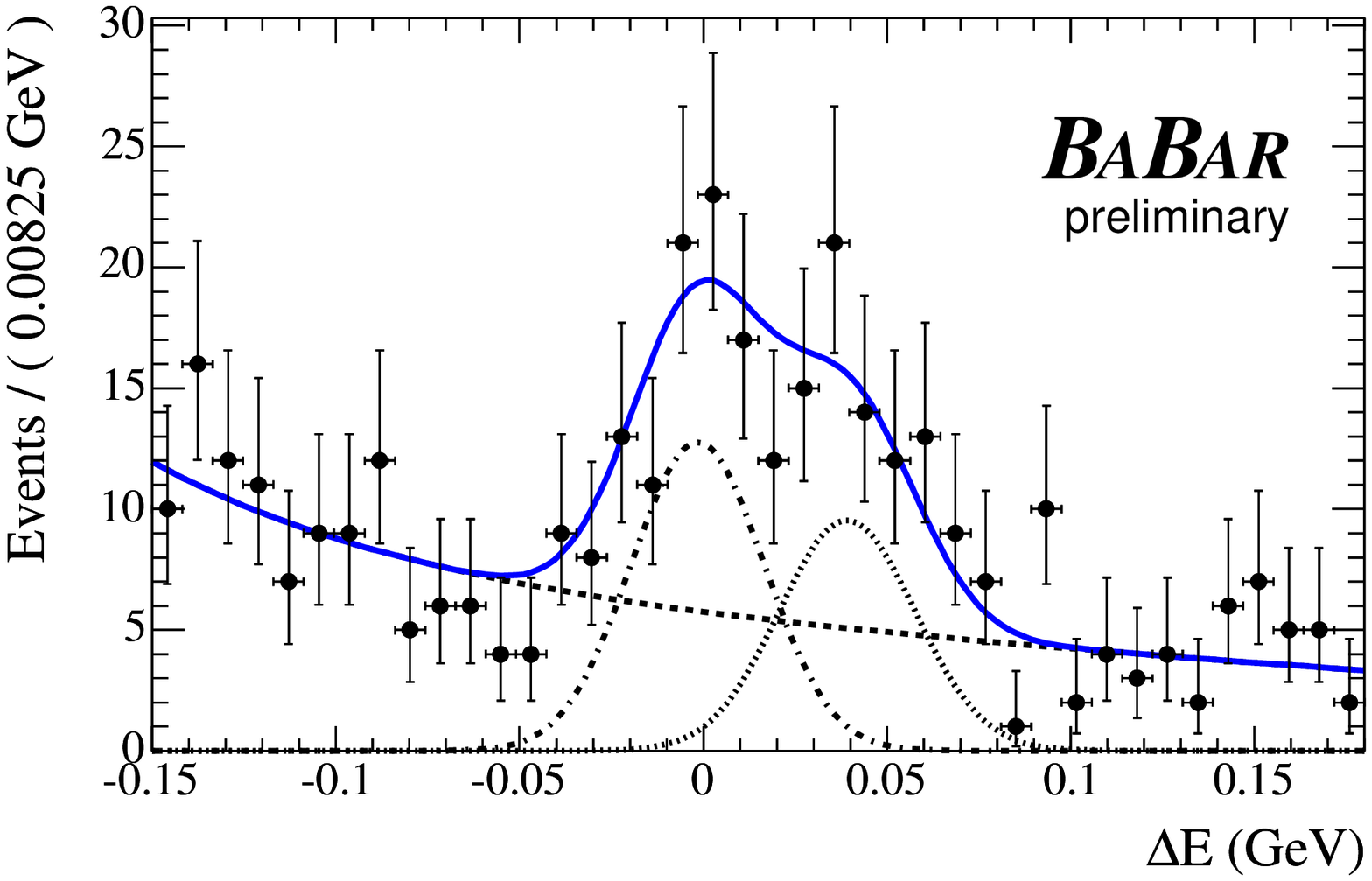}
      \caption{\DeltaE distributions of \btodh\ candidates, where a
	charged kaon mass hypothesis is assumed for $h$. Events are enhanced in
	$B^-\rightarrow D^0K^-$ purity by requiring the Cherenkov
	angle of the track $h$ to be within $2\sigma$ of the kaon
	hypothesis. Top: $B^-\ra D^0[K^-\pi^+]K^-$; middle: 
	$B^-\ra D^0_{\CPp}[\kk,\pipi]K^-$; bottom: $B^-\ra
	D^0_{\CPm}[\kspi0]K^-$. Solid curves 
	represent projections of the maximum likelihood fit; dashed-dotted,
	dotted and dashed curves represent the $B\rightarrow D^0K$,
	$B\rightarrow D^0\pi$ and background contributions.} 
      \label{fig:fit_kaons}
    \end{center}
  \end{figure}

  The results of the fit are summarized in Table~\ref{tab:fitresults}.
  Figure~\ref{fig:fit_kaons} shows the distributions of \DeltaE\ for the 
  $K^-\pi^+$, \CPp\ and \CPm\ modes after enhancing the $B\rightarrow
  D^0K$ purity by requiring that the prompt track be consistent with the
  kaon hypothesis. This requirement is about $95\%$ efficient for the
  \btodk signal while retaining only $4\%$ of the \btodp candidates.
  The projection of a likelihood fit, modified 
  to take into account the tighter selection criteria, is overlaid in
  the figure.
  
  \begin{table}[h]
    \caption{Results of the \btodk\ and \btodp\ yields from the maximum-likelihood fit on data.}
    \label{tab:fitresults}
    \begin{center}
      \begin{tabular}{lcccc}
	\hline
	\hline
	$D^0$ mode &\  $N(B\rightarrow D^0\pi)$ &\ $N(B\rightarrow
	D^0K)$ & $N(B^-\to \Dz K^-)$ & $N(B^+\to\Dzb K^+)$\\
	\hline
	$K^-\pi^+$      &\  $11930\pm 120$ &\ $897\pm 34$ & $441\pm 24$ & $456\pm 25$\\ 
	\hline
	$K^-K^+$        &\  $ 1093\pm  37$ &\ $ 75^{+13}_{-12}$ & $54^{+10}_{-9}$ & $22^{+8}_{-7}$\\
	$\pi^-\pi^+$    &\  $  345\pm  22$ &\ $ 18 \pm 7$ & $12\pm 5$ & $7^{+5}_{-4}$ \\
	\hline
	$K^0_S\pi^0$    &\  $ 1248\pm  40$ &\ $ 76^{+13}_{-12}$ & $46^{+10}_{-9}$ & $30^{+9}_{-8}$ \\
	\hline
	\hline
      \end{tabular}
    \end{center}

  \end{table}

  The double ratios $R_{\pm}$ are computed by scaling the ratios of
  the numbers of \btodk\ and \btodp\ mesons 
  by correction factors (ranging from 0.997 to 1.020 depending on the
  \Dz mode) that account for small differences in the efficiency 
  between the \btodk\ and \btodp\ selections, estimated with simulated signal 
  samples. The results are listed in Table~\ref{tab:final_ratio}. 

  The direct \CP\ asymmetries $A_{\CP\pm}$
  for the 
  $B^\pm\ra D^0_{\CP\pm}K^\pm$ 
  decays are calculated from the measured
  yields of positive and negative charged meson decays and the results
  are reported in Table~\ref{tab:final_ratio}.

  \begin{table}[h]
    \caption{Measured double branching fraction ratios $R_{\pm}$ and \CP
      asymmetries $A_{\CP\pm}$ for different \Dz\ decay modes. The first
      error is statistical, the second is systematic.} 
    \label{tab:final_ratio}
    \begin{center}
      \begin{tabular}{lcc}
	\hline
	\hline
	$D^0$ decay mode &\ \ \ \ \ \ $R_{\CP}/R$ &\ \ \ \ \ \ $A_{\CP}$\\
	\hline
	\hline
	$K^-K^+$ &\ \ \ \ \ \  $0.92\pm 0.16 \pm 0.07$&\ \ \ \ \ \ $0.43\pm 0.16 \pm 0.09$ \\
	$\pi^-\pi^+$ &\ \ \ \ \ \  $0.70\pm 0.29 \pm 0.09$&\ \ \ \ \ \ $0.27\pm 0.40 \pm 0.09$ \\
	\CP-even combined &\ \ \ \ \ \ $0.87\pm0.14\pm0.06$&\ \ \ \ \ \ $0.40\pm 0.15\pm 0.08$\\
	\hline
	$K^0_S\pi^0$ &\ \ \ \ \ \  $0.80\pm 0.14 \pm 0.08$&\ \ \ \ \ \ $0.21\pm 0.17 \pm 0.07$ \\
	\hline
	\hline
      \end{tabular}
    \end{center}
  \end{table}

  Systematic uncertainties in the double ratios $R_{\pm}$ and in the \CP 
  asymmetries $A_{\CP\pm}$ arise primarily from uncertainties in
  signal yields due to the estimate of the peaking backgrounds and
  from the imperfect knowledge of the PDF shapes. 
  The systematic uncertainty associated to peaking backgrounds is evaluated
  by taking into account the uncertainties on their branching fractions and 
  by allowing for Poisson fluctuations of their yields in the selected data
  sample. The estimated yields are $29\pm7$ ($B^-{\ra}K^-K^+K^-$),
  $4\pm4$ ($B^-{\ra}K^-\pi^+\pi^-$) and $0.0^{+5.6}_{-0.0}$ ($B^-\ra K^0_S \pi^0
  K^-$).
  Possible \CP asymmetries up to $30\%$ in their yields are also taken
  into account.
  The parameters of the PDFs that are fixed in the nominal fit are
  varied by $\pm1\sigma$ and the difference in the signal yields is
  taken as a systematic uncertainty.

  The uncertainties in the branching fractions of the channels 
  contributing to the \BB\ background have been taken into account.
  The correlations between the different sources of systematic errors, when
  non-negligible, are considered. An upper limit on intrinsic detector charge
  bias due to acceptance, tracking, and particle identification
  efficiency has been obtained from the measured
  asymmetries in the processes $B^-{\ra}D^0[{\ra}K^-\pi^+]h^-$ and 
  $B^-{\ra}D^0_{\CP\pm}\pi^-$, where \CP\ violation is expected to be
  negligible.
  This limit ($\pm 0.04$) has been added in quadrature to the total systematic
  uncertainty on the \CP\ asymmetry.

  \section{SUMMARY}
  \label{sec:Summary}
  In conclusion, we have reconstructed \btodk\ decays with
  $D^0$ mesons decaying to non-\CP (\kpi), \CP-even (\kk,\pipi) and
  \CP-odd (\kspi0) eigenstates. We have measured the \CP asymmetries
  $A_{\CPp} = 0.40\pm 0.15\stat\pm 0.08\syst$, $A_{\CPm} = 0.21\pm
  0.17\stat\pm 0.07\syst$, and the double ratio of branching fractions
  $R_{+} = 0.87\pm 0.14\stat\pm 0.06\syst$, $R_{-} =
      0.80\pm 0.14\stat\pm 0.08\syst$.
  These results improve the previous existing measurements from
  \babar~\cite{babarbtdk}.  All results presented in this document are
  preliminary.

  \section{ACKNOWLEDGMENTS}
  \label{sec:Acknowledgments}

  We are grateful for the 
extraordinary contributions of our \pep2\ colleagues in
achieving the excellent luminosity and machine conditions
that have made this work possible.
The success of this project also relies critically on the 
expertise and dedication of the computing organizations that 
support \babar.
The collaborating institutions wish to thank 
SLAC for its support and the kind hospitality extended to them. 
This work is supported by the
US Department of Energy
and National Science Foundation, the
Natural Sciences and Engineering Research Council (Canada),
Institute of High Energy Physics (China), the
Commissariat \`a l'Energie Atomique and
Institut National de Physique Nucl\'eaire et de Physique des Particules
(France), the
Bundesministerium f\"ur Bildung und Forschung and
Deutsche Forschungsgemeinschaft
(Germany), the
Istituto Nazionale di Fisica Nucleare (Italy),
the Foundation for Fundamental Research on Matter (The Netherlands),
the Research Council of Norway, the
Ministry of Science and Technology of the Russian Federation, and the
Particle Physics and Astronomy Research Council (United Kingdom). 
Individuals have received support from 
CONACyT (Mexico),
the A. P. Sloan Foundation, 
the Research Corporation,
and the Alexander von Humboldt Foundation.


\begin{thebibliography}{99}

  \bibitem{gronau1991}
    M.~Gronau and D.~Wyler, \jpl{B265}, 172 (1991); M.~Gronau
    and D.~London, \jpl{B253} 483 (1991); 
    D.~Atwood, I.~Dunietz and A.~Soni, \jprl{78}, 3257 (1997); 
    A.~Giri, Y.~Grossman, A.~Soffer, J.~Zupan \pr{D68} 054018 (2003). 

  \bibitem{detector}
    \babar\ Collaboration, B.\ Aubert {\em et al.}, 
    \nim{A479} 1 (2002).

  \bibitem{geant}
    GEANT4 Collaboration, S. Agostinelli {\em et al.}, \nim{A506} 250 (2003).

  \bibitem{PDG2004}
    Particle Data Group, S.~Eidelman {\em et al.}, \jpl{B592} 1 (2004).

  \bibitem{fox_wol}
    G.C. Fox and S. Wolfram, \jprl{41} 1581 (1978). 

  \bibitem{btokkk} Belle Collaboration, K.~Abe {\em et al.}, \pr{D65}
    092005 (2002); \babar\ Collaboration, B.\ Aubert {\em et al.},
    hep-ex/0308065, submitted to \jprl{}.

  \bibitem{babarbtdk} \babar\ Collaboration, B.~Aubert {\em et al.},
    \jprl{92} 202002 (2004).

  \end{thebibliography}
\end{document}